\documentclass[pra,twocolumn,10pt,aps]{revtex4-1}
\usepackage[utf8]{inputenc}  
\usepackage[T1]{fontenc}     
\usepackage[british]{babel}  
\usepackage[scaled=0.86]{berasans}  
\usepackage[colorlinks=true,allcolors=blue]{hyperref}  
\usepackage{graphicx} 
\usepackage[babel]{microtype}  
\usepackage{amsmath,amssymb,amsthm,bm,amsfonts,mathrsfs,bbm} 
\usepackage{setspace}
\usepackage{braket}
\usepackage[bottom]{footmisc}

\usepackage[bitstream-charter]{mathdesign}
\usepackage{tikz}
\usepackage{mathtools}
\usepackage{xspace}  
\usepackage{pgfplots}
\usepackage{verbatim}

\begin{document}

\title{Quantumness of Relative Incompatibility }

\author{Manish Kumar Shukla$^{1}$, Rounak Mundra$^{2}$, Arun K Pati$^{1}$,  Indranil Chakrabarty $^{1}$ $^{3}$, Junde Wu $^{4}$ }
\email[manish.shukla393@gmail.com,\\
rm29031996@gmail.com,\\
akpati@hri.res.in,\\
indranil.chakrabarty@iiit.ac.in,\\
wjd@zju.edu.cn.]{}
\affiliation{$^{1}$Quantum Information and Computation Group, Harish-Chandra Research Institute, Homi Bhabha National Institute, Allahabad 211019, India.\\
$^{2}$Center for Computational Natural Sciences and Bioinformatics$,$ International Institute of Information Technology$,$ Gachibowli$,$ Hyderabad$,$ India\\
$^{3}$Center for Security$,$ Theory and Algorithmic Research$,$ International Institute of Information Technology$,$ Gachibowli$,$ Hyderabad$,$ India.\\
$^{4}$ School of Mathematical Sciences, Zhejiang University, Hangzhou 310027, People’s Republic of China }

\begin{abstract}
We propose a new measure of relative incompatibility for a quantum system with respect to two non-commuting observables, and call it quantumness of relative incompatibility. In case of a classical state, order of observation is inconsequential, hence probability distribution of outcomes of any observable remains undisturbed. We define relative entropy of the two marginal probability distributions as a measure of quantumness in the state, which is revealed only in presence of two non-commuting observables. Like all other measures, we show that the proposed measure satisfies some basic axioms. Also, we find that this measure depicts complementarity with quantum coherence. The relation is more vivid when we choose one of the observables in such a way that its eigen basis matches with the basis in which the coherence is measured. Our result indicates that the quantumness in a single system is still an interesting question to explore and there can be an inherent feature of the state which manifests beyond the idea of quantum coherence.

\end{abstract}

\maketitle

\section{Introduction}

It is generally believed that quantum superposition and entanglement are two primordial quantum features, which are responsible for giving us significant advantages in the field of secure communication and computation over the digitized information system \cite{ekert1991quantum, bennett2014quantum, shor2000simple, adhikari2010probabilistic, bose1998multiparticle}.  However, these are not exhaustive features that are akin to quantum systems. Lately, relevant concepts like quantum correlations beyond entanglement \cite{ollivier2001quantum, henderson2001classical, modi2010unified, chakrabarty2011quantum}, coherence of quantum state \cite{winter2016operational}, set of stronger uncertainty relations \cite{maccone2014stronger, song2017stronger, wang2016experimental, mukhopadhyay2016stronger} and coherence of purification \cite{pati2019coherence} came into picture, where each of these concepts tell something about the quantum system. All these come under the umbrella of a universally accepted term called "quantumness". This motivates us to re-conceptualize the idea of what is quantum in a quantum state.\\

In particular, quantum coherence can be viewed as a fundamental signature of non-classicality in physical systems. It tries to capture  the superposition or wave aspect of a single quantum system \cite{winter2016operational, baumgratz2014quantifying, aberg2006quantifying}. Coherence can also be used as a resource for certain tasks like better cooling \cite{brask2015small, lostaglio2015description} or work extraction processes in nano-scale thermodynamics, many quantum algorithms \cite{anand2016coherence, shi2017coherence, olaya2008efficiency}, quantifying wave-particle duality \cite{bera2015duality, yuan2018experimental, abbott2008quantum} and in  biological processes \cite{bandyopadhyay2012quantum, wilde2009could}. The resource theory of quantum coherence \cite{streltsov2015measuring, asboth2005computable, xi2015quantum, streltsov2016entanglement, qi2017measuring, chin2017coherence, zhu2017operational, zhu2018axiomatic, korzekwa2016extraction} along with other resource theories of entanglement and thermodynamics \cite{hillery2016coherence, napoli2016robustness}, have also been established. Recently, a resource theory framework for the particle nature of a quantum system was introduced \cite{das2018quantifying} which quantifies what we mean by a 'particle' in quantum world. \\

Effort to understand and quantify the quantumness in a single pure quantum system is still not complete, and hence, it requires more investigation. Our idea is to quantify quantumness in a state as something that will depend on observables of our choice. Quantum coherence reveals one aspect of quantumness with respect to one observable (that fixes the basis). Here, in this article, we formalize the quantumness of the relative incompatibility of a state with respect to two non-commuting observables. Relative incompatibility is defined as relative entropy between the two marginal probability distribution of the outcomes of joint probability distributions. These joint probability distributions are obtained when measurements of the two non-commuting observables are performed in different order. We call this as the quantumness of relative incompatibility.  We show that our measure satisfies standard properties like convexity and additivity. We also show that for a pure quantum state, the quantumness shows complementary behaviour with the coherence of a system when we measure the coherence in the eigen basis of one of the observables. This indicates that quantumness may be fundamentally different concept than quantum coherence.\\ 

This article is organized as follows: section II begins by defining quantumness of a state. We show that it satisfies several realisable properties which are required to be qualified as a measure. In section III, we show the complementary nature of our measure with coherence, for a given choice of basis. In Section IV, we give certain examples of pure and mixed states to find out how quantumness behaves with state and observable parameters. Finally, we conclude in section V.

\section{Quantumness of a single quantum system}
In this paper, we introduce a novel way to characterize the quantumness of a single system. We try to quantify this through the deviation of a quantum system from its classical counterpart. In the quantum world, one of the key deviation from the classical feature is the non-commutativity of the observables. We define quantumness in the following subsection. \\ 

\textbf{Quantumness:} We select two non-commuting observables $A$ and $B$ and perform projective measurements $\Pi^{A}$ and $\Pi^{B}$ one after the other on the state $\rho$. This can be done in two different ways, i.e., $\rho \rightarrow \Pi^{A}(\rho) \rightarrow \Pi^{B}\Pi^{A}(\rho)$ or $\rho \rightarrow \Pi^{B}(\rho) \rightarrow \Pi^{A}\Pi^{B}(\rho)$ where $\Pi^{A}(\rho)$ = $\sum_{i} \pi_{i}^{A} \rho \pi_{i}^{A}$ with $\pi_{i}^{A} = |x_{i}\rangle \langle x_{i}|$ and $\Pi^{B}(\rho)$ = $\sum_{i} \pi_{i}^{B} \rho \pi_{i}^{B}$ with $\pi_{i}^{B} = |y_{i}\rangle \langle y_{i}|$. Now if we take marginal probability distributions $p^{ A}$ and $p^{\bold{'} A}$ of the observable say $A$ in each of these two cases ($p^{ A}$ is obtained when $A$ is measured first and $p^{\bold{'} A}$ when $A$ is measured after $B$), we find that the two distributions are different as long as the observables $A$ and $B$ are non-commuting. First of all, let us assume that measurement of observable $A$ involves projection on basis $\ket{x_1}$, $\ket{x_2}$, ..... $\ket{x_{n-1}}$ and $\ket{x_{n}}$. Similarly, for observable $B$, we have $\ket{y_1}$, $\ket{y_2}$, ...... $\ket{y_{n-1}}$ and $\ket{y_{n}}$ as the basis. Then we have, 
\begin{equation}
p^{ A} = \big[\langle x_{1} | \rho | x_{1} \rangle, \hdots \langle x_{i} | \rho | x_{i} \rangle, \hdots \langle x_{n} | \rho | x_{n} \rangle \big]^T. 
\label{eqn1}
\end{equation}
On the other hand, if we make a prior measurement of $B$, this changes the outcomes of $A$ in the following manner, $p_{j i}^{\bold{'}B A} = Tr [\Pi_{i}^{A}(\Pi_{j}^{B} \rho \Pi_{j}^{B})\Pi_{i}^{A}] $. After taking the marginal probability distribution we have,
\begin{equation}
p_i^{\bold{'} A}= \sum_{j} \langle y_{j} | \rho | y_{j} \rangle |\langle x_{i} | y_{j} \rangle|^2 ,
\end{equation}
for $i=1,2,...n$.
The quantumness $Q_{A,B}(\rho)$ for the  system $\rho$, under measurement of the non  commuting observables $A$ and $B$ is quantified by the relative entropy between these two marginal distributions $p^A$ and $p^{\bold{'} A}$. It is given by,
\begin{equation}
Q_{A,B}(\rho) = S(p^{ A}||p^{\bold{'} A}) , 
\end{equation}
where $S(.||.)$ is the relative entropy between the two probability distributions. The relative entropy or Kullback-Leibler divergence between two probability distributions $p(x)$ and $q(x)$ is defined as, $D(p||q)=\sum _{x\in X} p(x)\log \frac{p(x)}{q(x)}$. \\ 

\textbf{Note: } If we are looking at probability distribution of various outcomes of $A$ and we observe $B$ after observing $A$ as given in (Eq. \ref{eqn1}), it does not have any impact on the probability distribution of outcomes of $A$. The joint probability distribution is given by, $p_{i j}^{A B} = Tr [\Pi_{j}^{B}(\Pi_{i}^{A} \rho \Pi_{i}^{A})\Pi_{j}^{B}]$, where $\Pi_i^A$ and $\Pi_{j}^{B}$ are the projective measurements for $A$ and $B$ respectively. The marginal probability distribution of the observable $A$ is given by $p_{i}^{A}  = \sum_{j} Tr \Big[\Pi_{j}^{B}\Big(\Pi_{i}^{A} \rho \Pi_{i}^{A}\Big)\Pi_{j}^{B}\Big]= Tr \Big[\Big(\Pi_{i}^{A} \rho \Pi_{i}^{A}\Big)\sum_{j} \Pi_{j}^{B}\Pi_{j}^{B}\Big]=Tr \Big[\Big(\Pi_{i}^{A} \rho \Pi_{i}^{A}\Big)\Big]$. Hence, we see that the probability distribution of the outcomes of $A$ is unaffected if we observe $B$ after $A$. In fact, the probability distribution changes and the quantumness of the system is induced if we observe $B$ before $A$.\\

We now collect defining properties that any functional Q mapping states to the non negative real numbers, should satisfy in order for it to be a physically consistent measure.\\

\noindent \textbf{Q1:} Firstly, we require that the quantumness should be zero for all classical states, i.e., $Q_{A,B}(\rho)=0$ for all $\rho \in C$, where $C$ is the set of all classical states and $A$ and $B$ are the non-commuting observables.\\
\textbf{Q2:} Since the measure is defined through non-commuting observables, $Q_{A,B}(\rho)$ should be equal to zero for any state $\rho$ if observables $A$ and $B$ commute with each other. \\
\textbf{Q3:} (Convexity) From the physical perspective, we prefer that quantumness can only decrease under mixing. This leads to the convexity condition, i.e., $Q_{A,B}(\sum_i^n p_i \rho_i) \leq \sum_i^n p_i Q_{A,B}(\rho_i)$, where $p_i$ is the probability of mixing such that $\sum_{i} p_i=1$.\\
\textbf{Q4:} (Additivity) Given a measure $Q_{A,B}(\rho)$ and a state $\rho$ one may ask the question of what happens to the measure when we increase the number of states, in particular, is the measure additive or not. This is equivalent to the condition $Q_{A,B}(\rho^{\otimes n}) = n Q_{A,B}(\rho)$, that needs to be satisfied for all integer n. Any measure satisfying this property is said to be additive. A much stronger condition will be $Q_{A,B}(\rho_1 \otimes \rho_2)= Q_{A,B}(\rho_1)+Q_{A,B}(\rho_2)$.

\subsection{Classical System}
The measure proposed above is dependent on the observables $A$ and $B$, which means that a state $\rho$ can have zero quantumness for some specific values of $A$ and $B$ and non-zero for some others. On the other hand, it is also a measure of quantumness present in state with respect to two observables. This motivates us to ask: are there any states which give zero quantumness for all $A$ and $B$. We call such states as genuinely non-quantum states. These are similar to genuinely incoherent states, where coherence is zero irrespective of the basis we select. An example of single particle system with $Q_{A,B} = 0$ can be written as $\rho = \frac{1}{n} \sum_{k=1}^{n}\ket{\lambda_{k}}\bra{\lambda_{k}}$, where $\ket{\lambda_{k}}\bra{\lambda_{k}}$ is the eigen basis. The two marginals of the observable $A$ (for each of these cases when $A$ is measured before $B$ and $B$ is measured before $A$ ) are $p^{A} = p^{\bold{'}A}$ as $\big[\frac{1}{n},\frac{1}{n},.....,\frac{1}{n}]^T$ ($T$ is the transpose of the matrix).
Since $p_{i}^{A}$ and $p_{i}^{\bold{'}A}$ have the same probability distribution, quantumness $Q(\rho)$ (we will be using the notation $Q$ instead of using $Q_{A,B}$ from now on) comes out to be zero. 
\subsection{Commuting Observables}
Let us choose two commuting observables, $A$ and $B$. Since they are commuting, they will have same set of eigen basis. Let the orthogonal basis of observable $A$ be denoted by \{ $\ket{x_1}$, $\ket{x_2}$, ..... $\ket{x_{n-1}}$ and $\ket{x_{n}}$ \}. Basis of observable $B$ will just be the permutation of the basis of $A$. We denote the orthogonal basis of $B$ by \{$\ket{y_1}$, $\ket{y_2}$, ...... $\ket{y_{n-1}}$ and $\ket{y_{n}}$\}. Then $p^{A}$ can be written as, $p^{ A} = [\langle x_{i} | \rho | x_{i} \rangle] \ \ \ \text{for i = 1,2, ... n}$. 
Similarly $p^{\bold{'}A}$ can be expressed as, $p^{\bold{'}A} = [\sum_{i} \langle y_{i} | \rho | y_{i} \rangle \langle x_{j} | y_{i} \rangle|^2], \ \ \text{for j = 1,2,...n}$.

Now, since $ \langle x_{j} | y_{i} \rangle =\delta_{ij}$, we get $p^{A}$ equals to $p^{\bold{'}A}$. The relative entropy of two identical probability distribution is zero. So for commuting observables, quantumness $Q(\rho)$ of any state $\rho$ will be equal to zero. This clearly indicates this measure is a manifestation of the non-commutativity of two observables in quantum mechanics.   
\subsection{Convexity}
From a physical point of view, it is important to ensure that any measure which quantifies quantum behaviour in a system, should ideally reduce with mixing. This essentially means that the measure should be convex, i.e., $Q(\sum_i^n \omega_i \rho_i) \leq \sum_i^n \omega_i Q(\rho_i)$ (where $\omega_i$ is the probability of mixing). \\
Suppose that $\rho = \sum_{i}\omega_{i}\rho_{i}$, where $\omega_i$ are the associated weights with $\sum_{i} \omega_{i} = 1$. We take two observables $A$ and $B$ such that their measurement involves projection on '$k$' and '$n$' eigen basis, respectively. Then, $p_{kn}^{AB}$ is defined as, 
$p_{kn}^{AB} = Tr [\sum_{nk}(\Pi_{n}^{B}\Pi_{k}^{A})\rho(\Pi_{k}^{A}\Pi_{n}^{B})]$.  Using the above equation, $p_{kn}^{AB}$ can be transformed into $p_{kn}^{AB}  = Tr [\sum_{nk}(\Pi_{n}^{B}\Pi_{k}^{A})(\sum_{i} \omega_{i}\rho_{i})(\Pi_{k}^{A}\Pi_{n}^{B})] = \sum_{i} \omega_{i} Tr [\sum_{nk}(\Pi_{n}^{B}\Pi_{k}^{A})(\rho_{i})(\Pi_{k}^{A}\Pi_{n}^{B})]  = \sum_{i} \omega_{i} p_{kn}^{(i)AB}$. Similarly, we can write $p_{nk}^{BA}$ as $p_{nk}^{BA} = \sum_{i} \omega_{i} p_{nk}^{(i)BA}$.
Correspondingly, we can write $p_{k}^{A}$ and $p_{k}^{\bold{'}A}$ as $p_{k}^{A} = \sum_{i} \omega_{i} p_{k}^{(i)A} ,p_{k}^{\bold{'}A} = \sum_{i} \omega_{i} p_{k}^{\bold{'}(i)A}$.
Henceforth, the relative entropy of $p_{k}^{A}$ and $p_{k}^{\bold{'}A}$ becomes 
\begin{equation}
    \begin{split}
    S(p_{k}^{(i)A} || p_{k}^{\bold{'}(i)A}) & = S(\sum_{i} \omega_{i} p_{k}^{(i)A} ||                                                \sum_{i} \omega_{i}                                                                  p_{k}^{\bold{'}(i)A})\\
                                            & = (\sum_{i} \omega_{i}         p_{k}^{(i)A})\log(\frac{\sum_{i} \omega_{i} p_{k}^{(i)A}}{\sum_{i} \omega_{i} p_{k}^{\bold{'}(i)A}}). \\
    \end{split}
\end{equation}
Let X = $\sum_{i}\omega_{i}p_{k}^{(i)A}$ and Y = $\sum_{i}\omega_{i} p_{k}^{\bold{'}(i)A}$. Then the above equation boils down to $S(p_{k}^{(i)A} || p_{k}^{\bold{'}(i)A}) = X \log(X) - X \log(Y)$. This is a jointly convex function. Hence, 
$S(\sum_{i} \omega_{i} p_{k}^{A} || \sum_{i} \omega_{i} p_{k}^{\bold{'}A}) \leq \sum_{i} \omega_{i} S(p_{k}^{A} || p_{k}^{\bold{'}A})$.  It is a  very useful property and it captures the notion of the loss of information. It tells us that if we have  identifiable states $\rho_i$ that occurs with probability $\omega_i$, the weighted average of their individual quantumness undergoes a loss if we take a mixture of these states of the form $\sum_i^n \omega_i \rho_i$.

\subsection{Additivity}

Next, we see that whether our measure is additive or not, by which we mean that for any pair of states  $\rho_1$ and $\rho_2$, we have Q($\rho_1 \otimes \rho_2$)=Q($\rho_1$)+Q($\rho_2$) . The quantity $Q(\rho)$ also depends on the two observables $A$ and $B$. To collectively quantify the quantumness in more than one qubit system, we need to fix the observables $A$ and $B$, and then calculate the individual quantumness with respect to them. Let us assume that the measurement of observable $A$ involves two projective measurements on $\ket{x}, \ket{x_{\bot}}$ basis and the outcome of measurement $A$ on qubit $1$ is given by probability distribution, $ p^{A} =\big[ p_{1},(1 - p_{1})]^T$.
Similarly, the outcome of measurement $A$ on qubit $1$ after measurement of observable $B$ is given by,
$p^{\bold{'} A} = \big[ p_{1}^{'},(1 - p_{1}^{'})]^T$.
If we extend this to qubit $2$, we will similarly obtain two probability distribution functions corresponding to, first, the measurement of observable $A$ and second, the measurement of $A$ after measurement of $B$,
$p^{B} = \big[ p_{2},(1 - p_{2})]^T$ and $p^{\bold{'} B} = \big[p_{2}^{'},(1 - p_{2}^{'})]^T$, respectively.
Now, according to our definition of $Q(\rho)$, we have
Q($\rho_{1}$) = S($p^{A} || p^{\bold{'} A}$) and 
Q($\rho_{2}$) = S($p^{B} || p^{\bold{'} B}$). We define the joint quantumness as,
Q($\rho_{1} \otimes \rho_{2}$) = S($p^{A}p^{B} || p^{\bold{'} A}p^{\bold{'} B}$), where $ p^{A}p^{B} = \big[ p_1 p_{2},p_1(1-p_{2}),(1- p_1) p_{2},(1-p_1)(1 - p_{2})]^T$ and $ p^{\bold{'} A}p^{\bold{'} B}=\big[ p_1^{'} p_{2}^{'},p_1^{'} (1 - p_{2}^{'}),(1- p_1^{'}) p_{2}^{'}, (1-p_1^{'}) (1 - p_{2}^{'})]^T$, which is obtained by taking probabilities of all permutations of outcomes in two qubits. It can easily be shown  (see Appendix A) that the quantity $Q$ is additive in nature, i.e.,
\begin{equation}
    Q(\rho_{1} \otimes \rho_{2}) = Q(\rho_{1}) + Q(\rho_{2}).
\end{equation}

\section{Complementarity with coherence}
In the above section, we have defined the quantumness measure $Q(\rho)$ as the relative entropy of incompatibility for the  probability distribution of outcomes of an observable $A$ due to prior measurement of observable $B$. \color{black} There are other ways to quantify a quantum system such as coherence \cite{baumgratz2014quantifying}, entanglement, discord and many others. The most fundamental of these is coherence, which in some sense, gives a measure of the amount of superposition present in a quantum system. This motivates us to explore the connection between these two fundamental quantities that are trying to quantify the deviation from classical behaviour. The natural question is: are they similar or complementary in nature?  There are several measures of coherence like the $l_1$ norm, the relative entropy of coherence, etc. Here, we use the relative entropy of coherence $C(\rho)$ to establish a complementarity relation with $Q(\rho)$. The relative entropy of coherence is given by $C(\rho) = S(\rho_d) - S(\rho)$, where $S(\rho)$ is the Von-neuman entropy and $\rho_d$ is the density matrix obtained from $\rho$ by dropping all the off-diagonal elements. Since $Q(\rho)$ and $C(\rho)$ are both basis dependent quantities, we need to make several choices while trying to establish a relation between them.\\

\begin{figure}[h]
\begin{center}
\includegraphics[height=5.5cm,width=8.2cm]{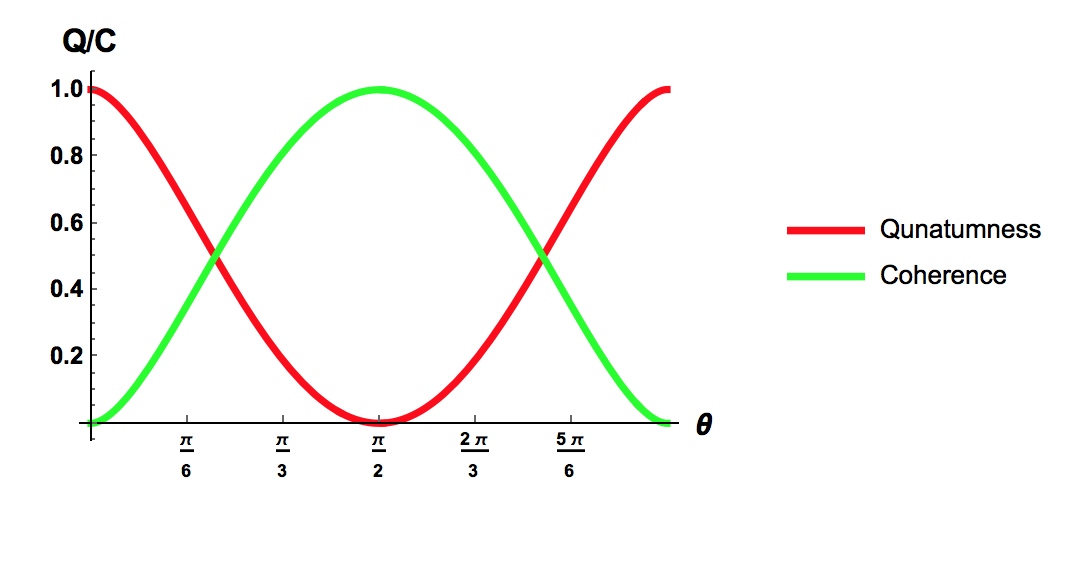}
\end{center}
\caption{ \scriptsize
The above figure shows the complementary behaviour of coherence and quantumness. The red line shows the value of quantumness and the green line shows the value of coherence in state $\ket{\psi} = cos(\frac{\theta}{2})\ket{0} +  sin(\frac{\theta}{2})\ket{1}$. The value of coherence is calculated in computational basis. The observable $A$ is fixed as measurement in computational basis and B as $\sigma_y$.}
\label{fig:complementarity}
\end{figure}
For a pure state $\rho$ we will always have $S(\rho) = 0$, which leaves us with $C(\rho) = S(\rho_d)$. If $\ket{\psi}$ is expressed in the basis $\ket{a_j}$, i.e., $\ket{\psi} = \sum_{i} c_i \ket{a_i}$, the coherence in $\ket{a_j}$ basis is given by -$\sum_i |c_i|^2 \log |c_i|^2$. \\

Now, we shift our attention towards calculation of $Q(\rho_{pure})$. As shown in the section above, $Q(\rho)$ is also a basis dependent quantity and depends on, (a) the basis in which $\rho_{pure}$ is expressed, (b) the eigen basis of observable $A$ and, (c) the eigen basis of observable $B$. Since we are making two measurements and computing relative entropy between the two probability distributions, we can safely fix  the basis ($|a_{i}\rangle$) in which the state $\rho_{pure}$ is represented. This leaves us with two places where we encounter dependence on basis. \\

\textbf{Q($\rho$) in given basis $\ket{a_j}$:}
To calculate $Q(\rho)$ in a given basis, we need to fix the observables $A$ and $B$ while calculating $p_{i}^{\bold{'} A}$ and $p_{i}^{A}$. Let us assume that the observable A has the eigen basis $\ket{\mu_i}$. Then $p_{i}^{ A}$ can be expressed as $p_{i}^{ A} = |\langle \psi | \mu_i \rangle |^2$.

Now, since we want to calculate $Q(\rho)$ in $\ket{a_i}$ basis, we choose observable $A$ to have basis $\ket{a_i}$, i.e., $\ket{\mu} = \ket{a}$. We maximize over the possible measurements of $B$ given by $\ket{b_i}$. The outcome after measurement of observable $B$ is given by, $\rho^{\bold{'}} = \sum_{i}^n |\langle \psi | b_i \rangle|^2 \ket{b_i}\bra{b_i}$. If we perform the measurement to observe $A$, we find that, $p_{i}^{\bold{'} A} = \sum_{j}^n |\langle \psi | b_j \rangle|^2 |\langle a_i \ket{b_j}|^2$.

By definition, we have $Q(\rho) = S(p_{i}^{\bold{'} A} || p_{i}^{ A})$, using which we can express
\begin{equation}
\begin{split}
    S(p_{i}^{ A} || p_{i}^{\bold{'} A}) = \sum_{i}^n x_i \log \frac{x_i}{z_i}\\
    = \sum_{i}^n x_i \log x_i - \sum_{i}^n x_i \log z_i,
    \end{split}
\end{equation}
where $z_i = \sum_{j}^n
    |
    \langle \psi | b_j \rangle |^2 |\langle a_i \ket{b_j}|^2$ and $x_i = |\langle a_i | \psi \rangle
    |^2$. 
Note that the first term is the negative of coherence of the pure state, i.e., $Q(\rho_{pure}) = -C(\rho_{pure}) - \sum_{i}^n x_i log z_i$. Since $Q$ is defined in terms of relative entropy, it will always be positive. Also, coherence in any state is always non negative. The maximum value being one, allowing us to claim that the second term, say $D = - \sum_{i}^n x_i log z_i $, is always greater than coherence, i.e., $- \sum_{i}^n x_i log z_i \geq C(\rho_{pure}) $. Thus $D$ is always a positive quantity, and for pure states we have $Q(\rho_{pure}) +C(\rho_{pure})=D$. If A and B are mutually unbiased, then 
$|\langle a_{i}| b_{j} \rangle |^2 = \frac{1}{d}$ and $D = log (d)$. In that case, we have
\begin{equation}
    Q(\rho_{pure}) + C(\rho_{pure}) = \log (d).
\end{equation}
This gives rise to complementary behaviour of quantumness $Q$ and coherence $C$, as the sum will always be bounded by a positive quantity $D$  (see figure (\ref{fig:complementarity}) as a particular example).
\\ 
 
\textbf{Note:} For a qubit in particular, if we use mutually unbiased basis as observable $B$, we will always have $p^{\bold{'} A} = \big[\frac{1}{2},\frac{1}{2}]^T$ and the quantity $D$  will always have value equal to one. $Q(\rho_{pure}) + C(\rho_{pure}) = 1$.  The marginal probability distribution ( $p^{A}$ ) of outcomes of A when we first observe A followed by B is,  $p^{A} = \begin{bmatrix}
        \langle 0 | \rho | 0 \rangle \\ 
        \langle 1 | \rho | 1 \rangle \\
        \end{bmatrix}
$. The marginal probability distribution ( $p^{\bold{'} A}$ ) of outcomes of A, when we make a prior measurement of B, which is in a mutually unbiased basis is, $p^{\bold{'} A} = \begin{bmatrix}
        \frac{1}{2} \\ 
        \frac{1}{2} \\
        \end{bmatrix}
$.
The quantumness of the qubit becomes, $Q(\rho_{pure}) = \ \ S(p^{A} || p^{\bold{'} A}) 
= \ \ \langle 0 | \rho | 0 \rangle \log\frac{\langle 0 | \rho | 0 \rangle}{p_{1}^{'}}  + \langle 1 | \rho | 1 \rangle \log\frac{\langle 1 | \rho | 1 \rangle}{1-p_{1}^{'}}= \langle 0 | \rho | 0 \rangle \log \langle 0 | \rho | 0 \rangle  + \langle 1 | \rho | 1 \rangle \log\langle 1 | \rho | 1 \rangle \\
        \ \ \ - \big[ \langle 0 | \rho | 0 \rangle \log p_{1}^{'} + \langle 1 | \rho | 1 \rangle \log (1 - p_{1}^{'})\big]  -C(\rho_{pure}) + \big[ \langle 0 | \rho | 0 \rangle  + \langle 1 | \rho | 1 \rangle \big]$.
Therefore, the coherence and quantumness satisfy $Q(\rho_{pure}) + C(\rho_{pure}) = 1$.
\section{Examples of Quantumness}
In this section, we take specific examples to study the behaviour of quantumness $Q$, by selecting various states and observables.
\subsection{Dependence on observable A and B}
\begin{figure}[h]
\begin{center}
\includegraphics[height=6.5cm,width=8.2cm]{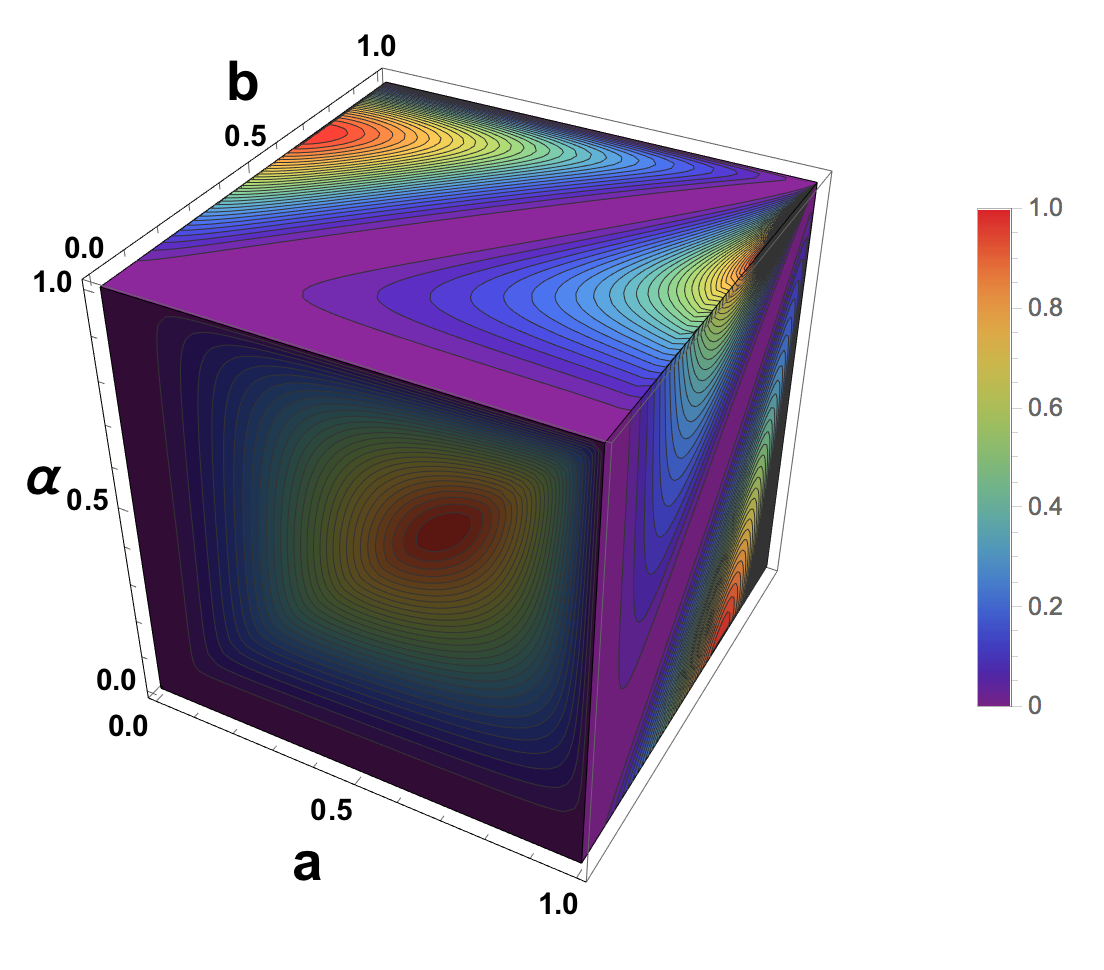}
\end{center}
\caption{\scriptsize
The above figure shows the variation in the value of $Q$ when we vary the state by changing the value of the state parameter $\alpha$ and vary the measurements $A$ and $B$, by changing $a$ and $b$ respectively.}
\label{fig:quantumness}
\end{figure}

Let us consider the case of state $\ket{\psi} = \alpha \ket{0} + \sqrt{1-\alpha^2}\ket{1}$ (where $\alpha$ is real). In addition to that we consider two observables $A$ and $B$ whose eigen states are given by $\ket{a} = a\ket{0} + \sqrt{1-a^2}\ket{1}$, $\ket{a_{\bot}} = \sqrt{1-a^2}\ket{0} - a\ket{1}$ and $\ket{b} = b\ket{0} + \sqrt{1-b^2}\ket{1}$, $\ket{b_{\bot}} = \sqrt{1-b^2}\ket{0} - b\ket{1}$ respectively. For simplicity we assume that $a,b$ and $\alpha$ take real values between $0$ to $1$. The relation between quantumness $Q$ in state $\ket{\psi}$ for observables A and B, as defined above, is shown in the contour plot given in figure (\ref{fig:quantumness}).
We notice that there are several high quantumness regions (where the color of the contour is red). The high quantumness of relative incompatibility region on the face where $b=0$, corresponds to observable A having this as basis \{$\ket{+}$, $\ket{-}$\} and observable B is fixed in computational basis \{$\ket{0}$, $\ket{1}$\}. We notice that for $\alpha^2 = \frac{1}{2}$ it attains the maximum value of $1$. Similarly, when we look at the other face where $a=1$ (which corresponds to measuring A in computational basis), we find that it is maximum for $\alpha = 1$ or $0$ and observable B has \{$\ket{+}$, $\ket{-}$\} basis. It takes minimum value when both observables are same. All the above examples show that quantumness in any state increases if we change the observables $A$ and $B$, such that measuring $B$ changes the outcome of $A$ to maximum uncertainty.

\subsection{Maximum quantumness in computational basis: pure state}
Let us consider a pure single qubit state given by, $\ket{\psi} = \cos(\frac{\theta}{2})\ket{0} + e^{i\phi}\sin(\frac{\theta}{2})\ket{1}$ (where $0 \leq \theta \leq \Pi$ and $0 \leq \phi \leq 2\Pi$). We want to quantify the quantumness of relative incompatibility in this state in computational basis. So we fix the observable $A$ as measurement in computational basis $\{\ket{0},\ket{1}\}$. We are free here to choose the observable $B$. If for any given state we choose the measurement $B$, which gives the maximum value of $Q$, we obtain the maximum quantumness in that state in the computational basis.  
\begin{figure}[h]
\begin{center}
\includegraphics[height=5.5cm,width=8.2cm]{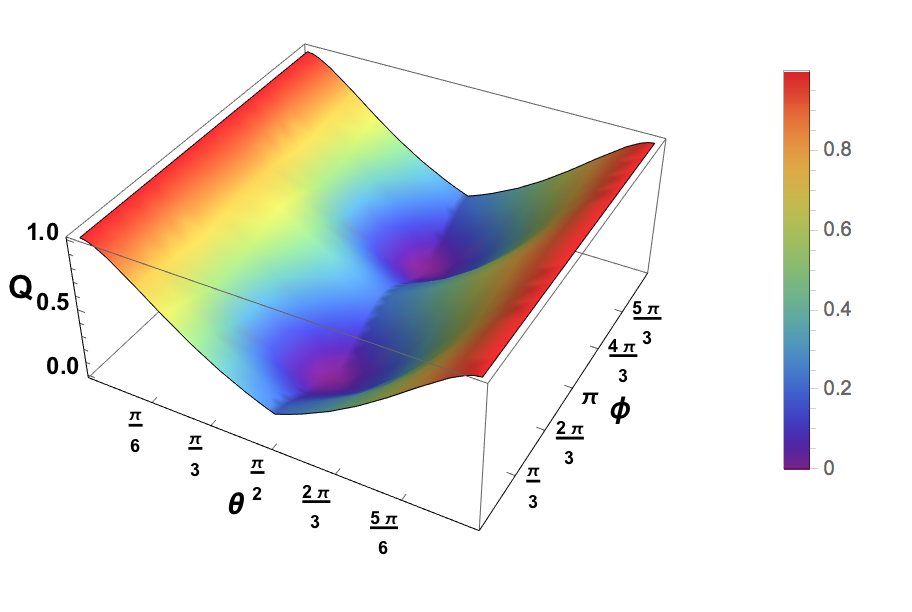}
\end{center}
\caption{\scriptsize
The above figure shows the value of maximum quantumness in general pure state, given by $\ket{\psi} = cos(\frac{\theta}{2})\ket{0} + e^{i\phi} sin(\frac{\theta}{2})\ket{1}$ (where $0 \leq \theta \leq \Pi$ and $0 \leq \phi \leq 2\Pi$) in computational basis \{$|0\rangle, |1\rangle$\}.}
\label{fig:maxquantumnessinpure}
\end{figure}
We have shown the value of maximum $Q$ ($Q_{max} = \max_{\substack{\ket{b_i}},} S(p^{ A}(\ket{b_1})||p^{\bold{'} A}(\ket{b_1}))$, where $\ket{b_i}$, is the basis set of an observable $B$) for this state in figure(\ref{fig:maxquantumnessinpure}). We see that the maximum value is obtained when the state $\ket{\psi}$ is close to either $\ket{0}$ or $\ket{1}$ and reduces symmetrically in the middle to obtain the minimum value when $\theta = \frac{\Pi}{2}$. The hue shows value of increasing quantumness ($Q$), where red denotes the maximum value and violet denotes the minimum. 

\subsection{Maximum quantumness in computational basis: mixed state}
We can extend the same analysis in computational basis to the mixed states. In this example we consider the state $\rho = p \ket{\psi}\bra{\psi} + (1-p) \frac{I}{2}$, where  $\ket{\psi} = cos(\frac{\theta}{2})\ket{0} + sin(\frac{\theta}{2})\ket{1}$ and $p$ is the mixing parameter. 
\begin{figure}[h]
\begin{center}
\includegraphics[height=5.5cm,width=8.2cm]{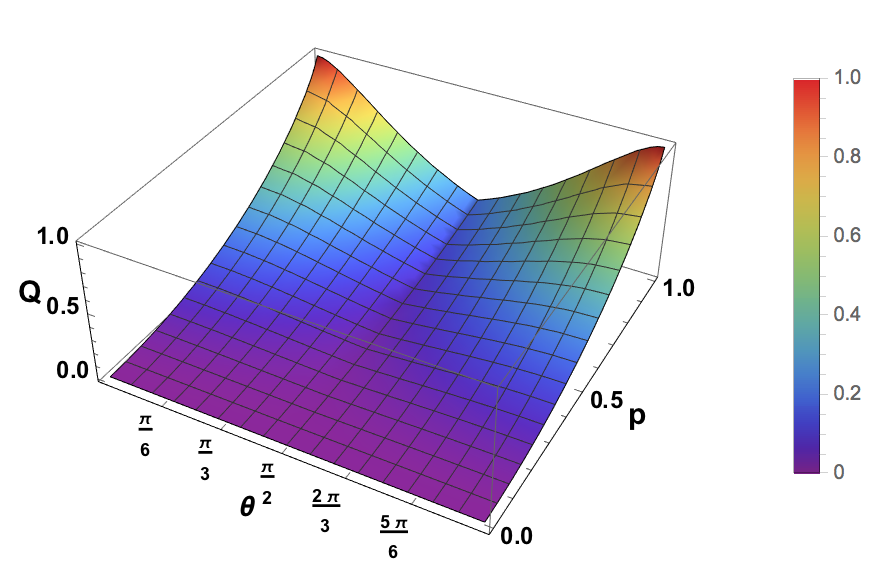}
\end{center}
\caption{\scriptsize
The above figure shows the value of maximum quantumness in mixed state of the form $\rho = p \ket{\psi}\bra{\psi} + \frac{1-p}{2}I$, where, $\ket{\psi} = cos(\frac{\theta}{2})\ket{0} +  sin(\frac{\theta}{2})\ket{1}$ in computational basis.}
\label{fig:maxquantumnessmixedstate}
\end{figure}
Figure \ref{fig:maxquantumnessmixedstate} shows the maximum value of $Q$ in the mixed state defined above. We see that with mixing, the value of $Q$ decreases. The maximum value is obtained when it is a pure state $\ket{\psi}$ and the minimum value is $0$, in case of completely classical state. We note a similar trend in pure states when we change the value of $\theta$, i.e., for $\theta = 0$ or $\Pi$ we obtain the maximum value for any given value of $p$ and the minimum value when $\theta = \frac{\Pi}{2}$. The hue shows value of increasing quantumness ($Q$) where red denotes the maximum value and violet denotes the minimum.\\

\section{Conclusion}
Quantifying the quantum nature and  understanding what is quantum in a single system is of fundamental importance. Here, we show that quantumness in a single system may depend on the choice of observables for which measurements are performed. In particular, we conceptualize quantumness of the relative incompatibility of a state with respect to two non-commuting observables. We have shown that it satisfies the standard properties like convexity and additivity. Most interestingly, for a given choice of basis of a particular observable, quantumness shows complementary feature with the coherence of a system. This clearly indicates that quantumness of a state is something quite different than the quantum coherence. It is something that is only revealed in presence of two non-commuting observables. We hope that our finding adds new aspects to the notion of quantum for a single system and may be useful in quantum information processing. \\ 

\noindent \textit{Acknowledgements}
This project is supported by the National Natural Science
Foundation of China (Grants No. 11571307,
and No. 61877054).


%

\appendix
\section{Additivity}

We take two qubit 1 and 2 and we assume that their states are represented by $\rho_{1}$ and $\rho_{2}$ respectively. The quantumness also depends on two observables A and B. We fix both the observables to quantify the quantumness in each qubit. The marginal probability distribution ( $p^{A}$ ) of outcomes of A on qubit 1 when we first observe A followed by B is given by, 
$
    p^{A} = \begin{bmatrix}
        p_{1} \\
        (1 - p_{1}) \\ 
        \end{bmatrix}
$
Similarly, the marginal probability distribution ( $p^{\bold{'} A}$ ) of outcome of A on qubit 1 when we first observe B followed by A is given by, 
$
    p^{\bold{'} A} = \begin{bmatrix}
        p_{1}^{'} \\ 
        (1 - p_{1}^{'}) \\
        \end{bmatrix}
$
Now we extend the same idea to qubit 2. The two marginal probability distribution ( $p^{B}$ and $p^{\bold{'} B}$ ) corresponding to outcome of A when A was observed before B and after B is given by,
$
    p^{B} = \begin{bmatrix}
        p_{2} \\ 
        (1 - p_{2}) \\
        \end{bmatrix}, \ \ \ 
        p^{\bold{'} B} = \begin{bmatrix}
        p_{2}^{'} \\ 
        (1 - p_{2}^{'}) \\
        \end{bmatrix}
$,
respecxtively. The quantumness of qubit 1 becomes,  
$Q(\rho_{1}) = p_{1}\log\frac{p_{1}}{p_{1}^{'}} + (1 - p_{1})\log\frac{(1 - p_{1})}{(1- p_{1}^{'})}$
Similarly, the quantumness of qubit 2 becomes, 
$Q(\rho_{2}) = p_{2}\log\frac{p_{2}}{p_{2}^{'}} + (1 - p_{2})\log\frac{(1 - p_{2})}{(1- p_{2}^{'})}$ 
The quantumness of composite system is given by, $Q(\rho_{1} \otimes \rho_{2})$ =  $S(p^{A}p^{B} || p^{\bold{'} A}p^{\bold{'} B})$. Here we have two independent joint probability distribution as, 
\begin{widetext}
$
p^{A}p^{B} = \begin{bmatrix}
        p_1 p_{2} \\ 
        p_1 (1 - p_{2}) \\
        (1- p_1) p_{2} \\ 
        (1-p_1) (1 - p_{2}) \\
        \end{bmatrix}, \ \ \ 
     p^{\bold{'} A}p^{\bold{'} B} = \begin{bmatrix}
        p_1^{'} p_{2}^{'} \\ 
        p_1^{'} (1 - p_{2}^{'}) \\
        (1- p_1^{'}) p_{2}^{'} \\ 
        (1-p_1^{'}) (1 - p_{2}^{'}) \\
        \end{bmatrix}
$
The quantumness of composite system becomes, 

\begin{equation*}
    \begin{split}
       Q(\rho_{1} \otimes \rho_{2}) &= \ \ p_{1}p_{2}\log\frac{p_{1}p_{2}}{p_{1}^{'}p_{2}^{'}} + (1-p_{1})(1-p_{2})\log\frac{(1-p_{1})(1-p_{2})}{(1-p_{1}^{'})(1-p_{2}^{'})} + p_{1}(1-p_{2})\log\frac{p_{1}(1-p_{2})}{p_{1}^{'}(1-p_{2}^{'})} + p_{2}(1-p_{1})\log\frac{p_{2}(1-p_{1})}{p_{2}^{'}(1-p_{1}^{'})} \\
       &= \ \ p_{1}p_{2}\big[\log\frac{p_{1}}{p_{1}^{'}} + \log\frac{p_{2}}{p_{2}^{'}}\big] + 
       (1-p_{1})(1-p_{2})\big[\log\frac{(1-p_{1})}{(1-p_{1}^{'})} + \log\frac{(1-p_{2})}{(1-p_{2}^{'})}\big] + p_{1}(1-p_{2})\big[\log\frac{p_{1}}{p_{1}^{'}} + \log\frac{(1-p_{2})}{1-p_{2}^{'}}\big] \\
       & \ \ \ \ \ \ \ \ \ \ p_{2}(1-p_{1})\big[\log\frac{p_{2}}{p_{2}^{'}} + \log\frac{(1-p_{1})}{1-p_{1}^{'}}\big] \\
       &= \ \ p_{1} p_{2} \log\frac{p_{1}}{p_{1}^{'}} + p_{1}(1-p_{2})\log\frac{p_{1}}{p_{1}^{'}} + p_{1}p_{2}\log\frac{p_{2}}{p_{2}^{'}} + p_{2}(1-p_{1})\log\frac{p_{2}}{p_{2}^{'}} + 
       (1-p_{1})(1-p_{2})\log\frac{(1-p_{1})}{1-p_{1}^{'}} \\
       & \ \ \ \ \ \ \ \ + p_{2}(1-p_{1})\log\frac{(1-p_{1})}{1-p_{1}^{'}} + (1-p_{1})(1-p_{2})\log\frac{(1-p_{2})}{1-p_{2}^{'}} + p_{1}(1-p_{2})\log\frac{(1-p_{2})}{1-p_{2}^{'}} \\
       &= \ \ p_{1}\log\frac{p_{1}}{p_{1}^{'}} + p_{2}\log\frac{p_{2}}{p_{2}^{'}} + 
       (1-p_{1})\log\frac{(1-p_{1})}{(1-p_{1}^{'})} + (1-p_{2})\log\frac{(1-p_{2})}{(1-p_{2}^{'})}\\
       &= \ \ Q(\rho_{1}) + Q(\rho_{2}) \\
    \end{split}
\end{equation*}
\end{widetext}


\end{document}